\documentstyle[mprocl]{article}

\def\Journal#1#2#3#4{{#1} {\bf #2}, #3 (#4)}

\def\PRD{{\em Phys. Rev.} D}

\def\CMP{\em Commun. Math. Phys.}

\def\be{\begin{equation}}
\def\ee{\end{equation}}
\def\bea{\begin{eqnarray}}
\def\eea{\end{eqnarray}}

\begin{document}

\title{CONSTRAINTS ON THE GEOMETRIES OF BLACK HOLES IN CLASSICAL AND 
	SEMICLASSICAL GRAVITY}

\author{ Paul R. Anderson and Courtney D. Mull}

\address{Department of Physics, Wake Forest University, \\ 
Winston-Salem, NC  27109, USA}

\maketitle\abstracts{
 Constraints on the geometries of static spherically
symmetric black holes are obtained by requiring that the spacetime
curvature be analytic at the event horizon.  Further constraints are
obtained by requiring that the semiclassical trace equation be satisfied
in the case that only conformally invariant fields are present.  It is
found that there exists a range of sizes for which zero temperature
black holes do not exist.  The range depends on the number and types
of quantized fields present.}
 
It is well known that the stress-energy of quantized fields can significantly alter 
the spacetime geometry near the event horizon of a black hole.  However, due
to the difficulties involved in computing the stress-energy of quantized fields
in black hole spacetimes, only the linearized semiclassical backreaction equations
have been solved.  Thus it is useful to try and see if
constraints can be placed on black hole solutions to the full nonlinear semiclassical
backreaction equations.  

Previously Mayo and Bekenstein \cite{MB}looked at
the problem of constraining static black hole solutions to Einstein's equations for various
types of matter fields.  They found that black hole solutions exist if the stress-energy tensor is finite on the horizon and one component satisfies a certain inequality.  The
geometry near the horizon is of the same form as the Schwarzschild geometry except in the
limit of an extreme black hole where the inequality becomes an equality.  In this latter case they put constraints on the form of the geometry near 
the horizon.  They also point out that if the transition between nonextreme 
metrics and an extreme metric is thought of as analogous to a phase transition, then it is likely that the metric of an extreme black hole
is of the same form near the horizon as the extreme Reissner-Nordstr\"{o}m black hole.  Otherwise
the transition would correspond to a third or higher order phase transition and such transitions
have not been observed in nature.  Recently Zaslavskii \cite{Z} has used a power series expansion of the metric to determine the
general form it takes near the horizons of near extreme and extreme charged
black holes in a cavity when the grand canonical ensemble is utilized.  

In this paper we take a different approach and first find constraints on the geometries
of static spherically symmetric black holes by imposing reasonable constraints on the
spacetime curvature at their event horizons.  The results are valid for any classical
or semiclassical metric theory of gravity.  We find that nonzero temperature
black holes must have geometries of the same form as the Schwarzschild geometry near their
event horizons.  Then we specialize to semiclassical gravity when only conformally invariant
fields are present.  By requiring that the trace of the semiclassical backreaction equations
be satisfied we find that there
exists a range of sizes for which zero temperature black holes cannot exist.  The size of
the range depends upon the number and types of quantized fields present. 

Our constraint on the curvature is to require that the curvature be analytic at the
event horizon in terms of the radial coordinate $r$, where the proper area of a two-sphere
centered on the origin is $4 \pi r^2$.  The assumption of analyticity,
while strong, is not without precedent.  Hawking \cite{H} assumed an analytic metric at the
event horizon in one of his uniqueness theorems for stationary rotating black holes.

To begin we write the metric for a static spherically symmetric spacetime in the general form
\begin{equation}
d s^2 = - f(r) dt^2 + \frac{1}{k(r)} dr^2 + r^2 d \Omega^2 \;\;\;.
\end{equation}
The unique non vanishing components of the Riemann curvature tensor in an 
orthonormal frame are
\begin{eqnarray}
R_{\hat{t} \hat{r} \hat{t} \hat{r}} &=&  \frac{v' k}{2} + \frac{v k'}{4} + \frac{v^2
 k}{4}  \\
R_{\hat{t} \hat{\theta} \hat{t} \hat{\theta}} &=& R_{\hat{t} \hat{\phi} \hat{t} \hat{\phi}}
  = \frac{v k}{2 r} \\
R_{\hat{r} \hat{\theta} \hat{r} \hat{\theta}} &=& R_{\hat{r} \hat{\phi} \hat{r} \hat{\phi}}
   = - \frac{k'}{2 r}  \\
R_{\hat{\theta} \hat{\phi} \hat{\theta} \hat{\phi}} &=& \frac{1 - k}{r^2}\;\;,
\end{eqnarray}
where $v \equiv f'/f$ and primes denote derivatives with respect to $r$.
The Kretschmann scalar $R_{\alpha\beta\gamma\delta} R^{\alpha\beta\gamma\delta}$
is proportional to the sum of the squares of these components.
If the spacetime has an event horizon then $f$ vanishes on that horizon.
The surface gravity at the event horizon is 
\begin{equation}
\kappa = \frac{v}{2} \left(f k \right)^{1/2} \;\;.
\end{equation}

We require that the above components of the Riemann tensor be analytic at the 
event horizon.  It is clear from Eq.(5) that $k$ must be analytic at the horizon.
  From Eq.(3) it is seen that the quantity $v\, k$ must also be analytic.  This second 
condition results in the further condition that $k \rightarrow 0$ at the 
event horizon.
To see this note that $v$ must diverge at the horizon because $v = f'/f$ 
and $f$
vanishes there.  These conditions can be summarized by saying that near 
the event
horizon $v$ and $k$ have the following leading order behaviors:
\begin{eqnarray}
v &=& p (r-r_0)^{m-n} \nonumber \\
k &=& q (r-r_0)^n  \;\;\;.
\end{eqnarray}
Here $p$ and $q$ are real positive constants, $m$ and $n$ are integers which 
satisfy the condition $n > m \ge 0$, and $r_0$ is the value of $r$ at 
the event horizon.

Further restrictions can be obtained from Eq.(2).  Substituting Eqs.(7)
into (2) one finds that the terms on the right hand side of (2) must either be separately 
finite at the  horizon or they must cancel.  They only cancel if $m = n-1$ and $p = 2 - n$.  
However we previously showed that $n \ge 1$.  Thus, since $p > 0$, the only case in which 
they cancel is $n = 1$,\, $m = 0$, \, $p = 1$.  In this case it is easy to see that 
near the horizon $f = c\, (r - r_0)$ for some $c>0$.  The surface gravity
is nonzero in this case so these solutions describe nonzero temperature black holes.
If the terms in (2) are separately finite then the restrictions are $m \ge 1$ and $2 m \ge n$.
Examination of Eq.(6) shows that such solutions describe zero temperature black
holes.

Semiclassical gravity can be used to place further constraints on the geometry 
in the case that only conformally invariant free quantized fields are present.  
In this case the trace of the semiclassical backreaction equations is
\begin{equation}
- R - 6 a \Box R = 8 \pi [\alpha \Box R + \beta (R_{\alpha\beta} R^{\alpha \beta} -
\frac{1}{3} R^2) + \gamma
       C_{\alpha\beta\gamma\delta} C^{\alpha\beta\gamma\delta}]
\end{equation}
where $a$ is the coefficient of an $R^2$ term in the gravitational Lagrangian and 
$\alpha$, $\beta$, and $\gamma$ are determined by the numbers
and types of quantized fields present \cite{BD}. 

If Eqs.(7) are substituted into Eq.(8) and the above conditions on $m$ and $n$ are
imposed, then the
leading order terms near the horizon can be computed for various values of $m$ and
$n$.  Consider first the case $2 m > n > m \ge 2$.  In the limit $r \rightarrow r_0$
Eq.(8) becomes
\begin{equation}
- \frac{16 \pi}{3 {r_0}^2}\, (\beta + 2 \gamma) - \frac{2}{{r_0}^2} = 0
\end{equation}
For all fields $\beta + 2 \gamma > 0$ \cite{BD}.  Thus
there are no solutions to the trace equation for values of $m$ and $n$ in this range.

The only other possibility is $2m = n \ge 2$.  
If Eqs.(7) are substituted into Eq.(8) for values of $m$ and $n$ in this range, the
limit $r \rightarrow r_0$ is taken, and the resulting equation is solved for $q$, the
result is 
\begin{equation}
q_{\pm}  =  \frac{3 {r_0}^2 - 32 \pi 
(\beta - \gamma) \pm (768 \pi^2 \beta^2 - 3072 \pi^2 \beta \gamma - 288 \pi 
\beta {r_0}^2 + 9 {r_0}^4 )^{1/2}}{4 \pi (\beta + 2 \gamma) p^2 {r_0}^2}   \;.
\end{equation}
Since the left hand side of (10) is positive and real, the right hand side must be also.  
However the right hand side is complex if $r_{-} < r_0 <r_{+}$ where
\begin{equation}
 r_{\pm} = 4 (\pi \beta)^{1/2} \left[1 \pm \left(\frac{2}{3 \beta} \right)^{1/2} (
\beta + 2 \gamma)^{1/2} \right]^{1/2} \;\;\;.
\end{equation}
For all allowed values of $\beta$ and $\gamma$\, $r_{+}$ is real.  If
$\beta < 4 \gamma$ then $r_{-}$ is imaginary and solutions only occur for
$r_0 > r_{+}$.  If $\beta > 4 \gamma$ then solutions also occur for $0 < r_0 < r_{-}
$.  Thus in all cases there is a range of values of $r$ for which zero temperature
black hole solutions to the semiclassical backreaction equations do not exist.

\section*{Acknowledgments}

This work was supported in part by Grant. No. PHY95-12686 from the
National Science Foundation.

\end{document}